\documentclass[balanced,aps,multicol]{revtex4}

 \usepackage{graphicx}
\begin{document}

\title{The collective brain is critical}

\author{Enzo Tagliazucchi$^{1}$\& Dante R. Chialvo$^{2,3,4}$}
\address{$^{1}$Departamento de F\'isica, Facultad de Ciencias Exactas y Naturales, Universidad de Buenos Aires, Argentina.
$^{2}$Consejo Nacional de Investigaciones Cient\'ificas y Tecnol\'ogicas (CONICET), Buenos Aires, Argentina. 
$^{3}$Facultad de Ciencias M\'edicas, Universidad Nacional de Rosario, Rosario, Argentina. 
$^{4}$David Geffen School of Medicine, UCLA, Los Angeles, CA. USA.}
 
\begin{abstract}
The unique dynamical features of the critical state can endow the brain with properties which are fundamental for
 adaptive behavior.  This proposal, put forward with Per Bak several years ago, is now supported by a wide body of 
empirical evidence at different scales demonstrating that the spatiotemporal brain dynamics  exhibits key signatures of
critical dynamics previously recognized in other complex systems. The rationale behind this program is discussed
in these notes, followed by an account of the most recent results, together with a discussion of the physiological 
significance of these ideas.
\end{abstract}

\maketitle

\section{Introduction:}

Throughout the years, brain theory  incarnated in various forms following contemporaneous technology. As expressed by  Braitenberg \cite{braitenberg}:
\emph{...Fascinating aspects of the latest technology are always creeping into science and being turned into subconscious motives for theory. For brain science, in the nineteenth century it was optics, with its idea of projection, and in the twentieth century it was radio engineering, with its logical circuits..."}. In the nineties, the fundamental concepts behind the physics of complex systems, motivated  us to work on ideas that now seem almost obvious: 1) the mind is a collective property emerging from the interaction of billions of agents; 2) animate behavior (human or otherwise) is inherently complex; 3) complexity and criticality are inseparable concepts.  These points were not chosen arbitrarily, but derived, as discussed at length here, from considering the dynamics of systems near the critical point of a order-disorder phase transition. Simply put, this view considered the brain as \emph{just another} dynamical system at criticality, knowingly non unique.
 
The brain seen as a dynamical object as discussed in these notes is grounded on accessible empirical evidence: as brain activity unfolds in time at all spatial scales, different patterns evolve. Proper measurements provide quantitative information regarding these patterns (for example, neuron spike trains, local field potentials, metabolic signals, behavioral measures, etc). The question is whether is it possible to explain all these results from a single fundamental principle, as it is the tradition in physics. And, in case the answer is affirmative, what does this unified explanation of brain activity  implies about goal oriented behavior? We will submit that, to a large extent, the problem of the dynamical regime at which the brain operates it is already solved in the context of critical phenomena and phase transitions. Indeed several fundamental aspects of  brain phenomenology have an intriguing counterpart with dynamics seen in other systems when posed at the edge of a second order phase transition. 

The paper is organized as follows: first, emergent complex phenomena and the intimate connection between criticality and complexity will be described. This will be in Section II, which reviews earlier work on ant's swarm, a metaphorical mind in itself, which is mathematically very close to nowadays models of collective decision-making. The large scale analysis of brain dynamics will be introduced in Section III, where the experimental results indicating criticality are discussed in detail. Section IV provides examples of how complexity and criticality manifest themselves in a relatively smaller scale (scale of a few millions of neurons). As mentioned above, the statistical properties of the brain dynamics must be reflected in the dynamics of  behavior. This is treated in Section V where recent reports seems to show such correspondence. Section VI, dwells into the evolutionary argument by which the brain and  the animate behavior must be critical anytime an organism needs to survive and evolve in an environment which, by thermodynamic reasons, is also critical. Secton VII is dedicated to discuss the tendency in the field to consider equilibrium models driven by external noise to accommodate the empirically observed fluctuations. Section VIII closes the paper with a prospective of some relevant issues to pursue further.

\section{Emergent complex dynamics is always critical}

Emergence refers to the observation of dynamics that is not expected from the systems equations of motion and, almost by (circular) definition, is  exhibited by complex systems.  As discussed at length elsewhere \cite{bakbook, maya, chialvo2004,chialvo2007,chialvo2010,sole, turcotte}, three features are present in complex systems: (I) they are  \emph{large} conglomerate  of \emph{interacting} agents, (II) each agent own dynamics exhibits some degree of \emph{nonlinearity} and (III) energy enters the system. \footnote{As an example, think that the agents are humans who are driven by food, sunlight and other energy sources, they can form families and communities. After many generations, eventually some political structure arises. Whatever the type of structure that emerges is unlikely to appear in the absence of any of the three elements emphasized above.}
These three components are necessary for a system to be able to exhibit, at some point, emergent behavior. It is well established that a number of isolated linear elements cannot produce unexpected behavior (mathematically, this is the case in which all motion can be formally predicted).

%%%%%%%%%%%FIGURE 1
\begin{figure}[htbp]
\centering{\includegraphics[width=0.3\textwidth,clip=true]{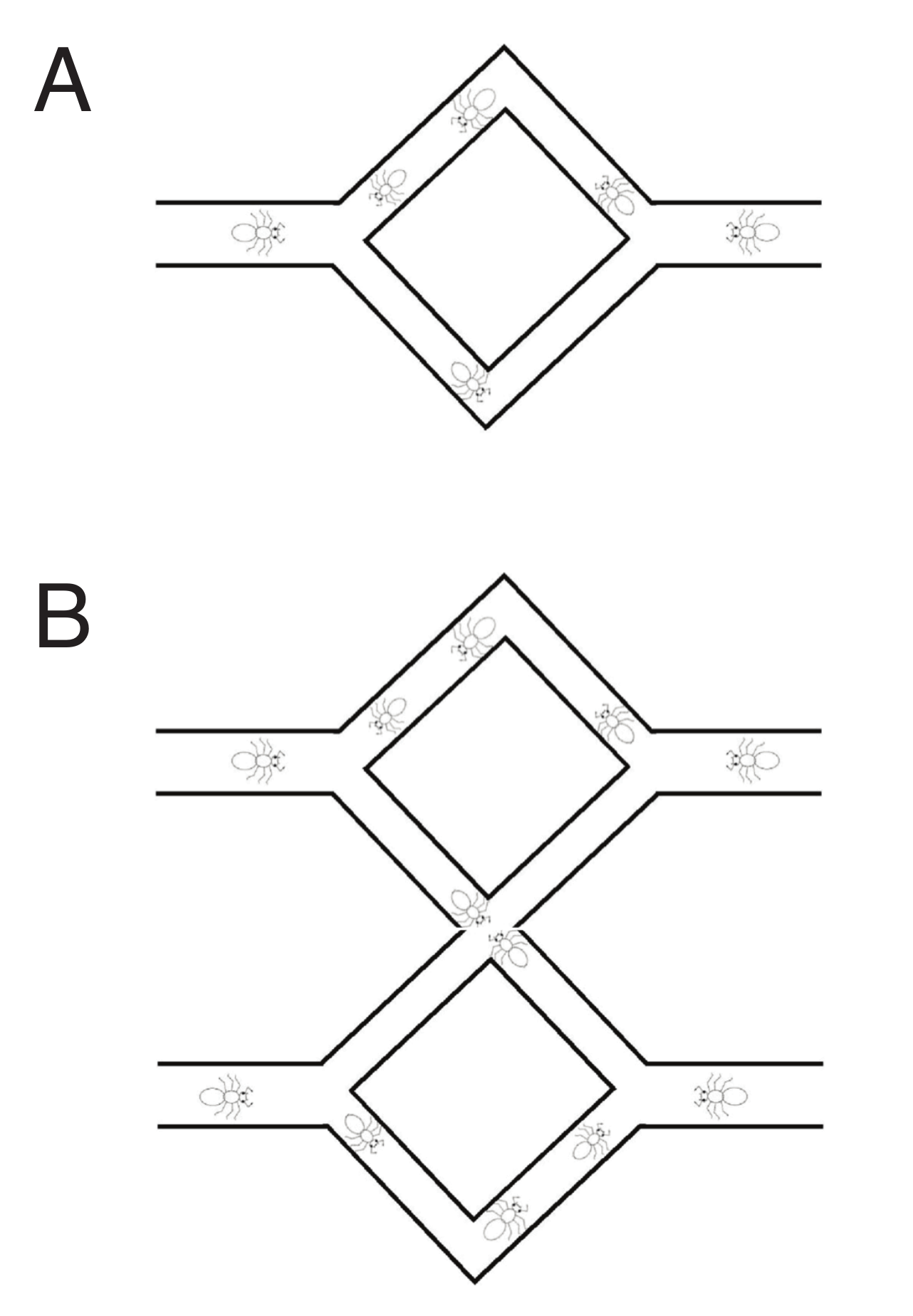}}
\caption{A: In typical binary bridge experiments, ants are allowed to walk over bridges connecting two or more areas with different concentration of pheromones. The dynamics of decision-making changes depending on the pheromone lying in the surface and on the ant's density. B: Gedanken experiment: the beginning of a network constructed by connecting binary bridges.}
\label{puentecitos}
\end{figure}

An inspiring example of emergence of complexity is the dynamic of swarms \cite{millonas1,millonas2,millonas3}, which we used  in the past as a toy model to understand how cognition could arise in neural networks \cite{rauch,chialvomillonas}. Of course, the case for social insects representing a paradigm of cooperative dynamics was extensively discussed before \cite{wilson71}. The uncertainty was centered at what collective property of foraging ants allows the emergence of trails connecting the nest with the food sources, forming structures spanning sizes several order of magnitude larger than any of the individual's temporal or spatial scales.  
To clarify that, Millonas introduced \cite{millonas1} a spatially extended model of what he called ``protoswarm''. His objective was  to understand the microscopic mechanism by which relatively unsophisticated ants can build and maintain these very large macroscopic structures. \footnote{Most readers will recognize the similarities with  brain's phenomenology.}

In the swarm model, there are two variables of interest, the organisms (in large number) and a field, representing the spacial concentration of a scent (such as pheromone). As in real ants, the model's organisms are influenced in their actions by the scent field and in turn they are able to modify it by depositing a small amount of scent in each step. The scent is slowly evaporating as well. The organisms only interact through the scent's field. The model is inspired in the behavior of real ants, in which they are exposed to bridges connecting two or more areas where
the ants move, feed, explore, etc. Eventually they will discover and cross one of the bridges. As it is illustrated in Figure \ref{puentecitos}A they will come to a junctions where they have to choose again a new branch, and continue moving. Since ants both lay and follow scent as they walk, the flow of ants on
the bridges typically changes as time passes. In the example illustrated in Figure 1A, after a while
most of the traffic will eventually concentrate on one of the two branches. The collective switch
to one branch is the emergent behavior, something that can be understood intuitively on the basis
of the positive feedback between scent following, traffic, and scent laying. What is less obvious is linking these ideas with the microscopic  rules  governing the dynamics of a single ant.  Numerous mathematical models and computer simulations were able to capture the ants behavior in the bridge experiment. \cite{beckers,deneubourg1,pasteels}.

%%%%%%%%%%%%%%%%%%%%%%%%%%%%%%%%%%%%%%%%%%%%%%%FIGURE 2 %%%%%%%%%%%%%%
\begin{figure}[htbp]
 \centering{\includegraphics[width=0.6\textwidth,clip=true]{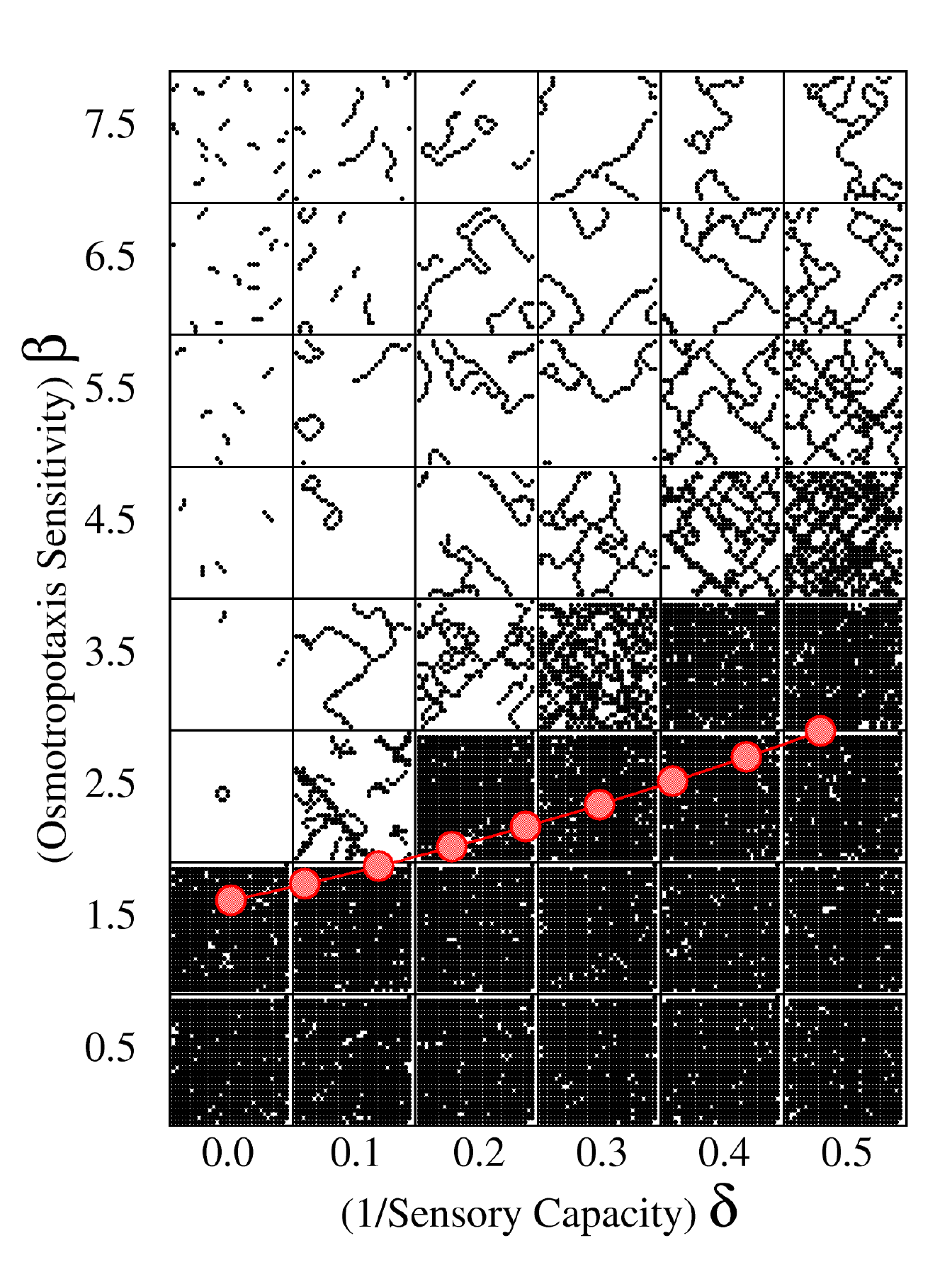}}
\caption{Disordered, complex and ordered patterns can be observed as asymptotic dynamic in
 synthetic swarms for different values of gain ($\beta$)
and dynamic range ($\delta$) in the ants' sensory system. For each parameter $\beta$ and $\delta$
value indicated on the axis the dots in each of the 48 squares represent a snapshot of the ants'
positions at the last ten iterations (after 5000 time steps starting from a random initial
distribution of agents). The circles joined by a line are the values for the phase
transition predicted by the theory \cite{millonas1,millonas2,millonas3}. There are three distinct behaviors: one region (below the line) where ants trajectories are
random, eventually covering the entire arena; a second dynamics (near and above the line) where lines of traffic over complex paths appear, and yet another patterns with clusters of immobile ants (far above the line). The arena is a 32x32 lattice and the other parameters as in \cite{rauch,chialvomillonas} . Replotted from  \cite{chialvo2008} .}
\label{antphase}
\end{figure}
%%%%%%%%%%%%%%%%%%%%%%%%%%%%%%%%%%%%%%%%%%%%%%%%%%%%%%%%%%%%%%%%

The bridge experiment results can be used as building blocks to more sophisticated settings, such as ants freely exploring an extended arena. The insight comes from a rather clever way that Millonas's model discretized space. The model can be viewed, for descriptive purposes, as a network constructed by connecting each point of a square lattice to its eight nearest neighbors, as in the cartoon of Figure \ref{puentecitos}B.  Thus, at each step each ant makes a decision to choose one of eight bridges; and deposits a fixed amount of pheromone as it walks. The decision is based on the amount of scent at each of the eight locations. The ants' sensory apparatus embedded in a physiological response function was modeled following biological realism, having two parameters: one which could be considered analogous to gain and the other  the inverse of sensory capacity (or dynamic range). The plot in Figure \ref{antphase} condenses the results from  the numerical simulations with different values for the physiological response function. At each combination of the explored $\beta$ and $\delta$ values there is a square plot which depicts the locations of each ant at the last ten steps of the simulation. It can be seen that ants converge to different behaviors depending on the parameter' values. For values of both small gain and small dynamic range  ants execute a random path, resulting in the plots fully covered, as in the right bottom corner. Thus, because of the low sensitivity ants are just making random choices at each juncture. For large enough gain (top left corner), ants senses saturate resulting in clusters of immobile ants in the same attracting spot. It is  between these two states, one disordered and the other frozen, that the swarm can organize and maintain large structures of traffic flow as those seen in nature.

%%%%%%%%%%%%FIGURE 3%%%%%%%%%%%%%%%
\begin{figure}[htbp]
 \centering{\includegraphics[width=0.5\textwidth,clip=true]{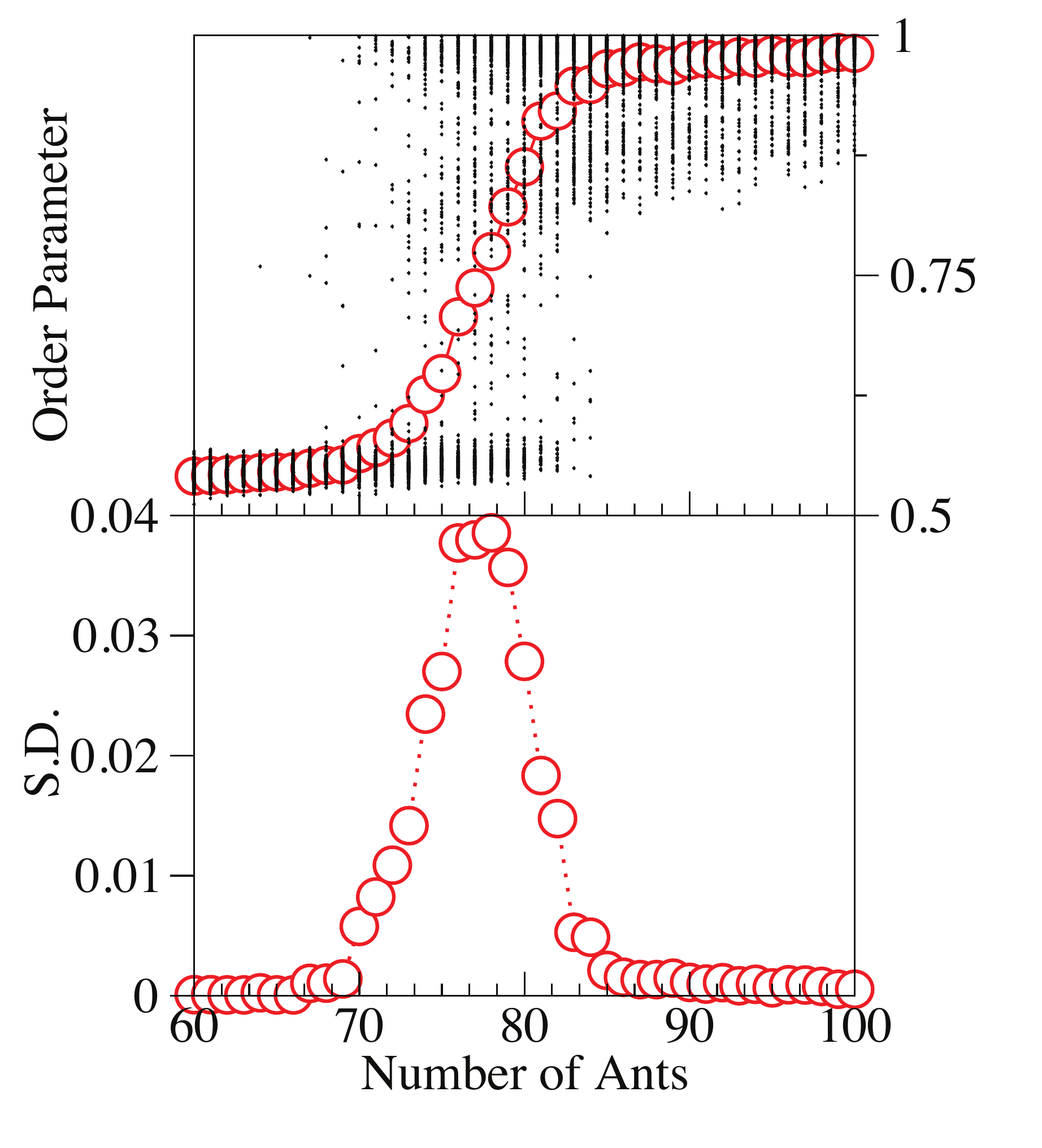}}
\caption{Collective swarm behavior near the phase transition. Collective order (top) increases  as a
function of increasing density of ants (i.e., the pheromone field is more frequently renewed). Note the generic increase in the amplitude of the fluctuations of the order parameter (bottom) near the critical point. Order parameter values from individual runs are plotted using dots and circles for
the average. S.D: Standard deviation of the data in the top panel. Replotted from \cite{chialvo2008}.}
\label{transition}
\end{figure}
%%%%%%%%%%%%%%%%%%%%%%%%%%%%%%%%%%%%%%%%%%%%%%%%%%%%%%%%%%%%%%%%

The stability analysis  in \cite{millonas1} makes straightforward to understand the transition between the disordered walks and the complex structures of trails (line and circles in Figure \ref{antphase}). Intuition already suggests that the model's $\beta$,$\delta$ values at which the order-disorder transition happens must depend on the number of ants able to reinforce the scent field. In Figure \ref{transition} results from several hundreds runs with fixed parameters and increasing density of ants are shown. The degree of collective order was evaluated by the proportion of ants ``walking'' on lattice points having above average scent concentration (see further details in \cite{rauch,chialvomillonas}). The top panel shows a plot of the results where the dots indicate the outcome of each individual run (with different initial conditions) and
circles the average of all runs. For low ants' density the expected random behavior is observed,
with equal likelihood for ants to be in or out of a high scent field. For increasing number of ants
the swarm suddenly starts to order until it reaches the point in which the majority of the ants are walking on a
field with scent concentration larger than the average.  Notice that, as the density approaches the transition, the amplitude of the order parameter's fluctuation increases (see bottom panel of Figure 3), a divergence which is generic for criticality. Thus, at the transition, large trial-to-trial variability is found for repeating realizations of the same numerical experiment. The most significant point here is to note that the best performance for the swarm, namely where long trails are formed (see Figure 2) corresponds also to the conditions for maximum variability. Thus maximum variability and performance coexist. 
\footnote{The brain correlate for this divergence will be a large variance in some order parameter (extracted from the dynamics) coexisting with optimal response to the same input, for instance.}  

These results were the first to show the simplest (local, memoryless, homogeneous and isotropic)
model which leads to trail forming, where the formation of trails and networks of ant traffic is
not imposed by any special boundary conditions, lattice topology, or additional behavioral rules.
The required behavioral elements are stochastic, nonlinear response of an ant to the scent, and a
directional bias. There are other relevant properties, discussed in detail elsewhere
\cite{rauch,chialvomillonas}, that arise \emph{only} at the region of ordered line of traffic,
including the ability to reconstitute trails  and amplify weak traces of scent, in analogy to memory traces. The conclusions important to these notes are that simple local rules allow the emergence of complex self organized patterns, which extend in a non local way beyond the scale of a single agent and which appear suddenly in a dynamical regime between random and uniform behavior, the ``critical state''.

The remaining of these notes will show evidence of an analogous behavior when neurons interact nonlinearly in the human brain, giving rise to the emergence of complex non-local patterns, as well as the functional and behavioral consequences of these patterns.

\section{The collective large-scale brain dynamics}

To visualize the collective activity of the individual constituents of the human brain (neurons) is harder, of course, than to contemplate the ants of the previous sections. For this purpose different methodologies have been developed which are capable to register this activity at different timescales and with different limitations. Large scale activity (resulting from the averaging of thousands of neurons) can be measured with excellent spatial and good temporal resolution using fMRI (functional magnetic resonance imaging). This non-invasive technique  allows indirect measurements due to metabolic and oxygenation changes correlated with synaptic activity, finally encoded in the BOLD   (Blood Oxygen Level Dependent) signal.

Patterns of global activity can be analyzed studying the relationship between the behavior of different members of the collective, in this case, regions of cortical tissue on the millimeter scale. Functional interactions between these areas can be studied comparing their BOLD signal time-courses. These functional interactions are not only present when the brain engages in a task but also during rest (i.e. spontaneous activity, in contrast to evoked activity). During the remainder of this notes we will focus mostly on resting state brain dynamics. These interactions can be represented using the concept of a network or a graph \cite{eguiluz,sporns2004,chialvo2004,sporns2010,salvador, vanden}. A graph consists of nodes connected by links, in this case the nodes are voxels (the smallest cubic regions that the spatial resolution of the method can resolve) and links represent coordinated activity between these two linked voxels. 

\begin{figure}[htbp]
\centering{\includegraphics[width=0.5\textwidth,clip=true]{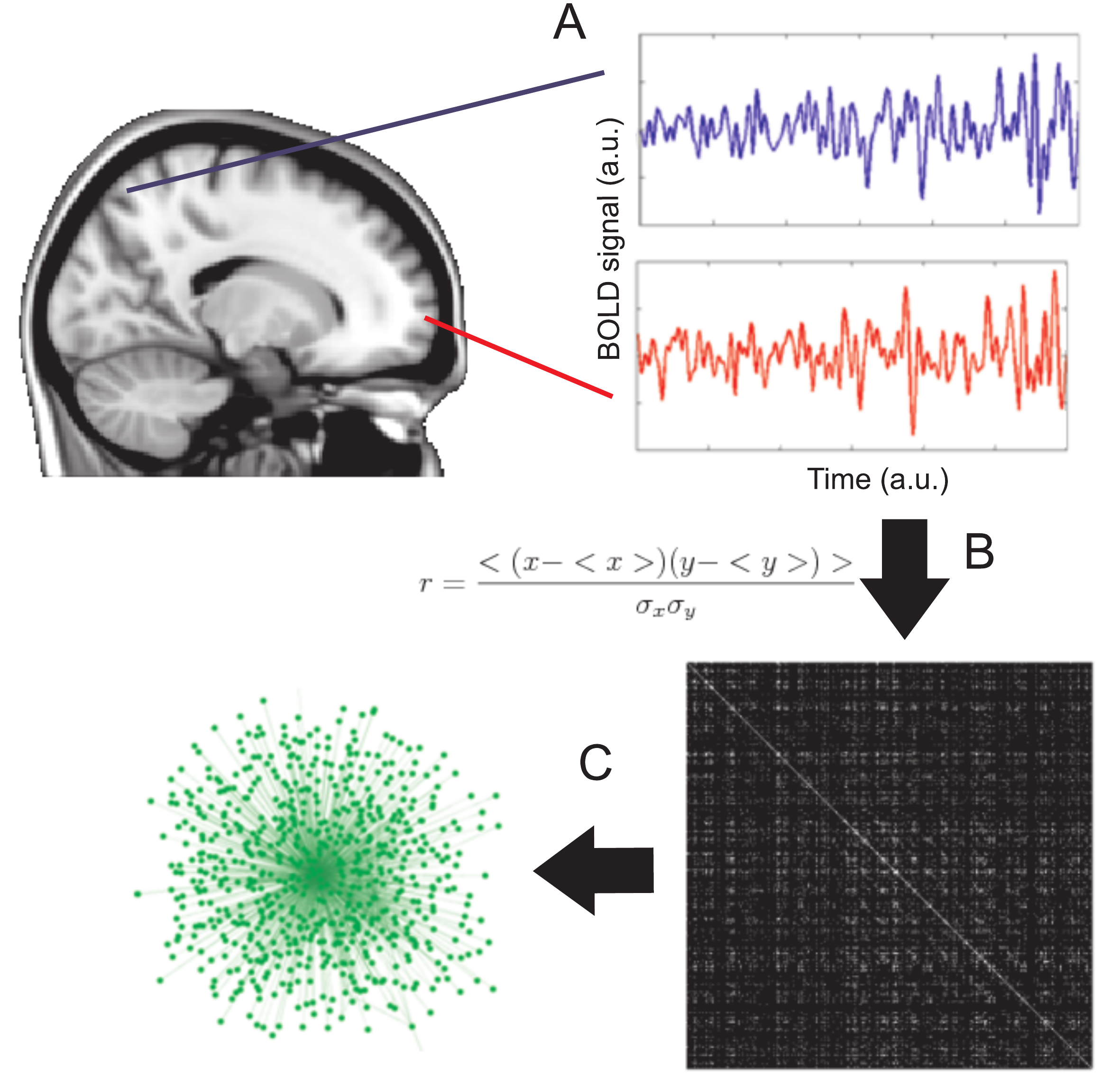}}
\caption{Procedure to extract brain functional connectivity networks. A: BOLD signal is extracted for every pair of voxels (in this case, voxels in the precuneus and middle frontal gyrus). B: Linear correlation (r) is computed between each pair of signals, leading to a correlation matrix. C: Correlations are thresholded at a given arbitrary value, a link is established between two nodes (or voxels) if the correlation of their respective BOLD signals exceeds the threshold, leading to a graph or network.}
 
\label{procedure}
\end{figure}

Different measures of coordination between the time course of the BOLD signal may be employed, the simplest being linear correlation. Functional connectivity networks can be constructed computing the linear correlation between BOLD signal pairwise for all voxels and introducing a link between voxels if they are correlated beyond an arbitrary threshold (for a diagram of the procedure see Figure \ref{procedure}). For a wide and reasonable range of thresholds, functional brain networks thus constructed are scale free, this is, the degree histogram of the voxels follows a power law of exponent approximately 2 (where degree is defined as number of neighbors in the network) \cite{eguiluz,vanden}. Also, functional connectivity networks have the small world property \cite{achard, basset2006a, basset2006b,stam} (the mean number of links that must be crossed to connect a given pair of nodes  is small relative to the local structure of the network) and are assortative (connectivity is stronger between nodes of similar degree). These properties are shared with a large number of networks constructed from different biological and social systems \cite{sporns2004,sporns2010}. They have strong implications regarding information transfer, stability and general robustness of the networks. Together, the scale free and small world properties imply the presence of long range (non-local) functional interactions between brain regions.
%%%%%%%%%%%%%%%FIGURE 4 %%%%%%%%%%%%
%%%%%%%%%%%%%%%%%%%%%%%%%%%%%%
 Most important to these notes, functional connectivity networks constructed from time-courses extracted from a paradigmatic example of a system with critical dynamics (Ising model for ferromagnetic materials) are virtually indistinguishable from brain functional connectivity networks \cite{fraiman}. This striking similarity between both networks is not imposed in the model: it arises as a very general consequence of critical dynamics, thus giving an explanation to the topological properties of brain functional connectivity networks solely based on the concept of criticality. 
%%%%%%%%%%%FIGURE 5 %%%%%%%%%%%
\begin{figure}[htbp]
\centering{\includegraphics[width=0.8\textwidth,clip=true]{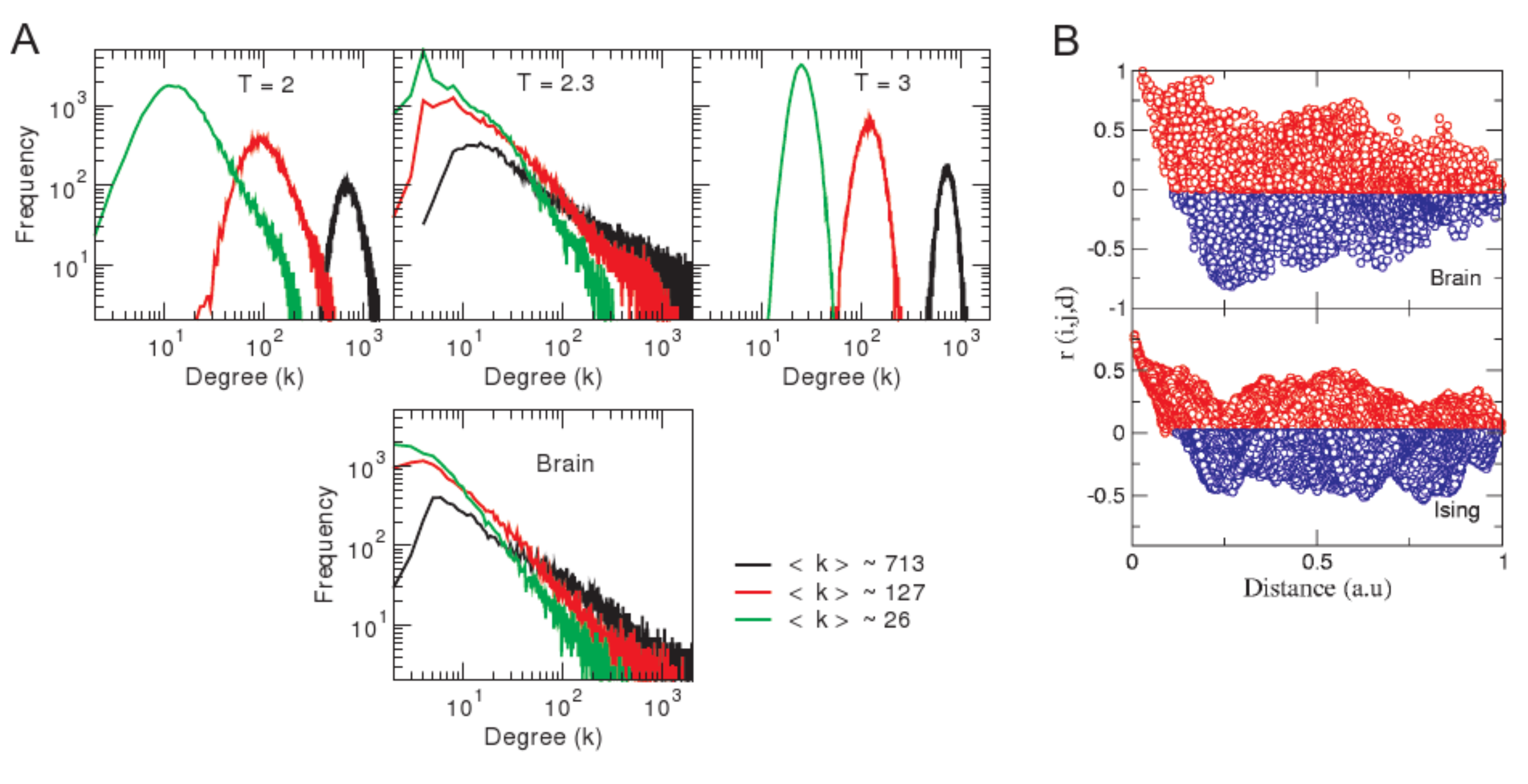}}
\caption{A: Degree distribution of functional connectivity networks extracted from the Ising model at different temperatures and different mean degree ($\left\langle k \right\rangle$)  (up) and degree distributions of brain functional connectivity networks (down) B: Correlation vs. distance to a given node for brain functional connectivity networks (up) and for Ising model networks (down). Adapted from \cite{fraiman} }
\label{ising}
\end{figure}
 
Functional connectivity networks are a useful global description of the brain, albeit a static one. They can be regarded as the ``skeletons'' of the dynamical systems, over which patterns of activity unfold over time. Deeper insight is gained from dynamical analysis (temporal and spatiotemporal) of large scale brain activity. Imaging experiments have gathered evidence showing that (even in the absence of explicit perception and cognitive performance) this activity displays complex behavior. In the purely temporal domain,  it is not homogeneous nor periodic (as EEG, MEG, fMRI, etc, recordings indicate) neither there is a representative or predominant frequency, as shown by the  1/f decay of spectral density. 
 
In the spatiotemporal domain, fMRI whole brain recordings reveal how activity explores complex patterns with long range correlations and anticorrelations \cite{fox2006,fox2007}. Strikingly, these patterns can be reduced to a small number of prototypical ones using a mathematical technique termed Independent Component Analysis (ICA) \cite{beckmann2004,beckmann2005}. Careful examination  shows that  these structures appear and disappear over time in the fashion of ``passing clouds'' that last only a few seconds. It has been shown that the spatial maps of these resting state independent components (also called resting state networks or RSNs) can be associated with different cognitive functions and are strongly correspondent with activation maps obtained during task performance \cite{smith}.

\begin{figure}[htbp]
\centering{\includegraphics[width=0.8\textwidth,clip=true]{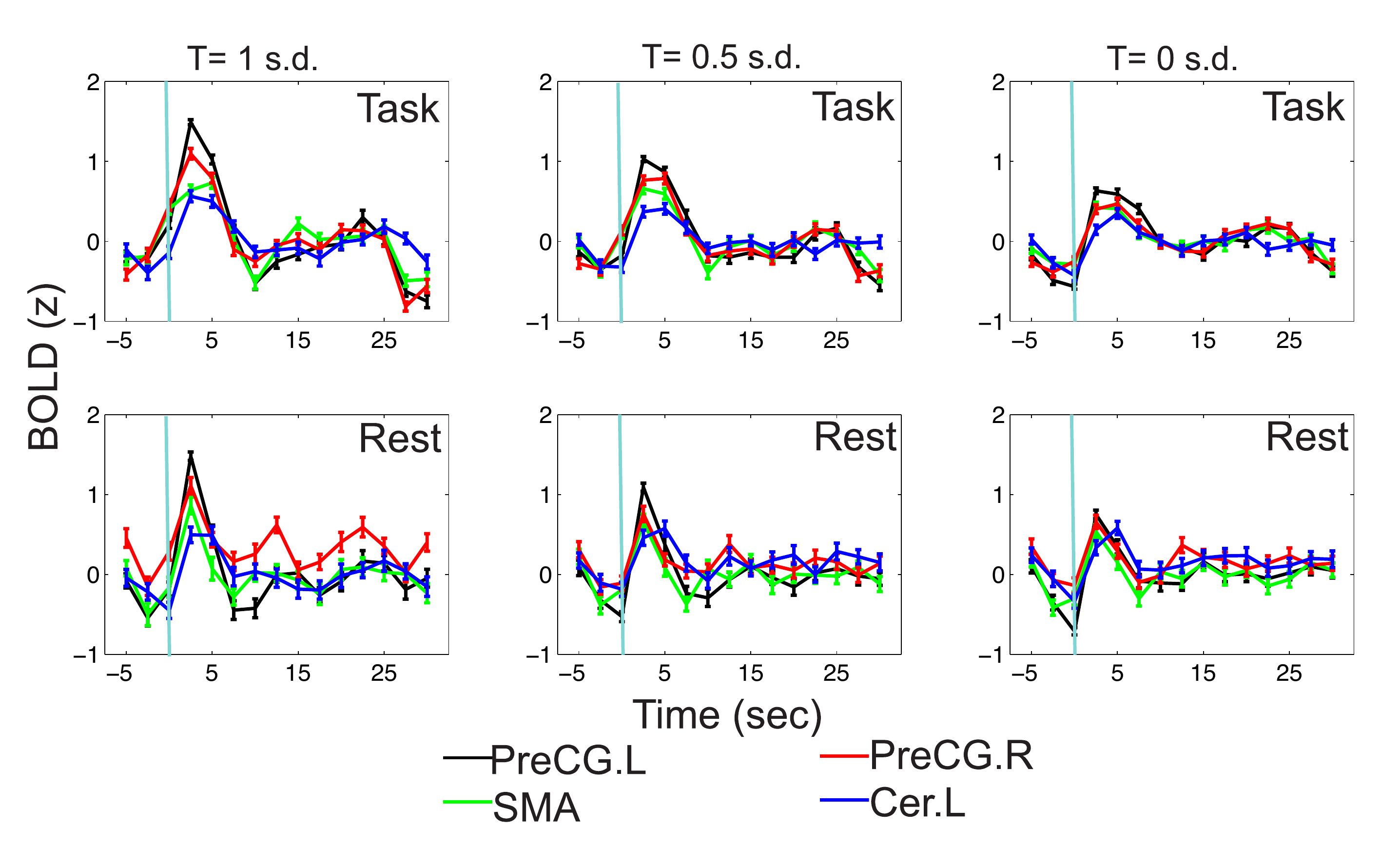}}
\caption{Averages of the BOLD signal triggered by increments of activity (beyond 0.5 S.D. from the mean) in left primary motor cortex  (PreCG.L) during right finger tapping (up) and resting state (down). Whenever the signal at PreCG.L exceeds the threshold, 30 sec. of the signal are extracted from right primary motor cortex (PreCG.R), supplementary motor area (SMA) and left cerebellum (Cer.L) and averaged. Notice the similarity between task and rest conditions, not only in shape but also in BOLD intensity. Replotted from \cite{enzo} }
\label{rbeta}
\end{figure}

This correspondence supports a dynamical view of resting state, in which activity continuously explores patterns (which are ``metastable states'') only to get temporarily locked in one of them as a task is executed or a sensory system stimulated. Not surprisingly, it is  when operating at the critical point that a dynamical system becomes most efficient in the \emph{flexible} exploration of these metastable states. This dynamical view of resting state need to be exploited given the strong  support \cite{enzo} from recent experiments involving simple motor tasks and resting state.  Briefly, it has been shown that average BOLD signals associated with spontaneous co-activations of key motor cortical areas strongly resemble co-activations during evoked activity, not only in location and temporal shape but also in intensity. In other words, spontaneous activations are completely analogous to evoked activations, except that the first ones are transient and their timing is uncertain. This correspondence is shown in Figure \ref{rbeta} for regions in the primary motor cortex, supplementary motor area and cerebellum, during finger tapping and rest. The continuous exploration of these patterns during resting state may have strong neurophysiological relevance, yet to be fully understood.

%%%%%%%%%%%%% FIGURE 6 %%%%%%%%%%%

It must be noted that while resting-state and spontaneous brain activity seems an ill defined concept, the bulk of results described in this last paragraph has not only been observed in conscious behaving humans but also during sleep and anesthesia. A great consistency is observed across subjects: approximately 10 minutes of fMRI measurements from a single subject are enough to reveal the main resting state networks (as shown in Figure \ref{dantenature})

\begin{figure}[htbp]
\centering{\includegraphics[width=0.45\textwidth,clip=true]{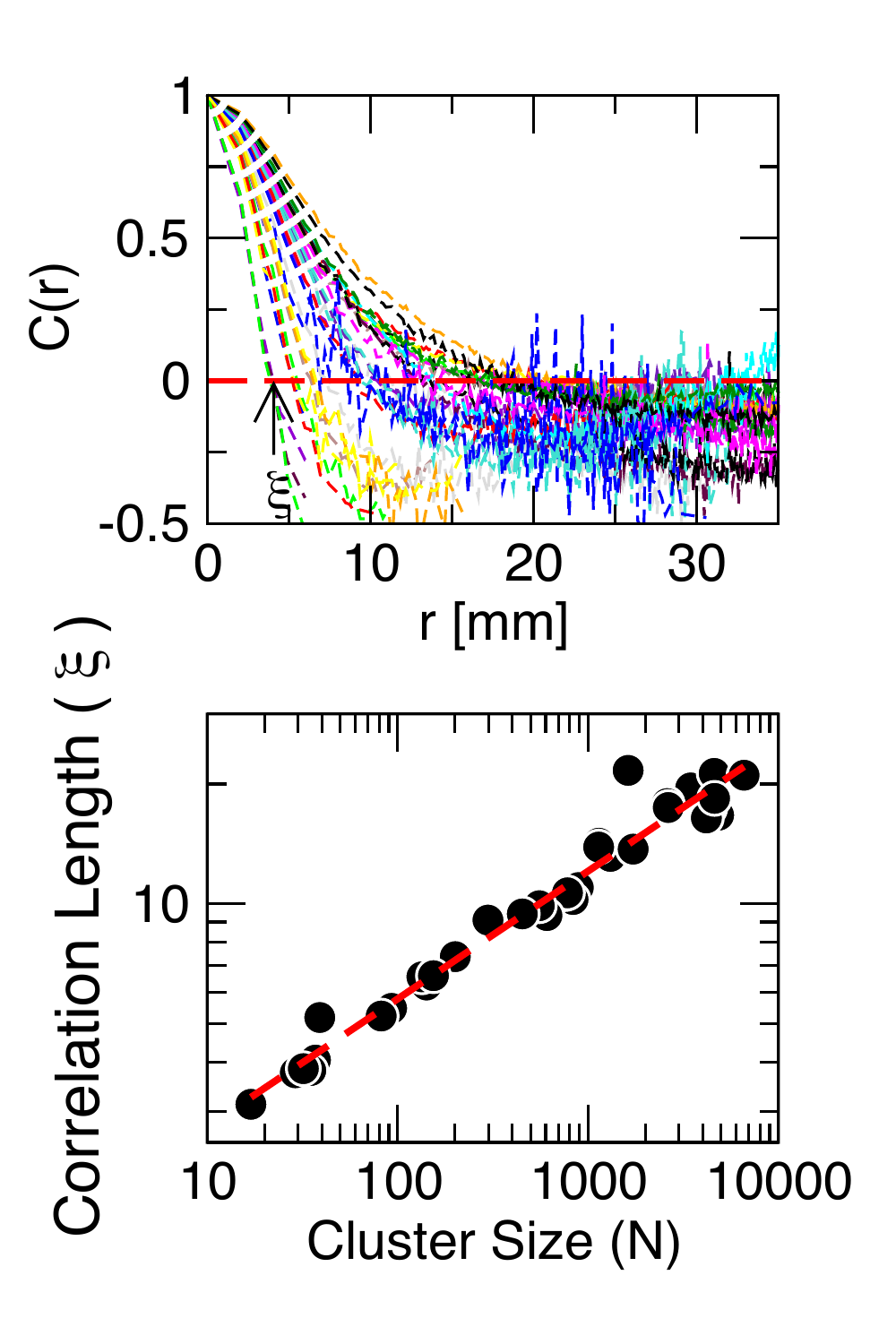}}
\caption{Top panel: increase of correlation length with cluster size. Each line shows the cross-correlation of BOLD activity fluctuations as a function of distance averaged over all time series of thirty five clusters (fragments of resting state networks). Bottom panel: Functional dependence of the correlation length (crossing of each curve of top panel through zero). The correlation length grows linearly with the cluster diameter. Thus, for example, the BOLD activity of two voxels spaced 4 mm apart on a cluster composed by 20 voxels, are as strongly correlated as two points 10 mm apart on a cluster composed by 1000 voxels. Replotted from \cite{chialvofraiman}.}
\label{dani}
\end{figure}

\begin{figure}[htbp]
\centering{\includegraphics[width=0.9\textwidth,clip=true]{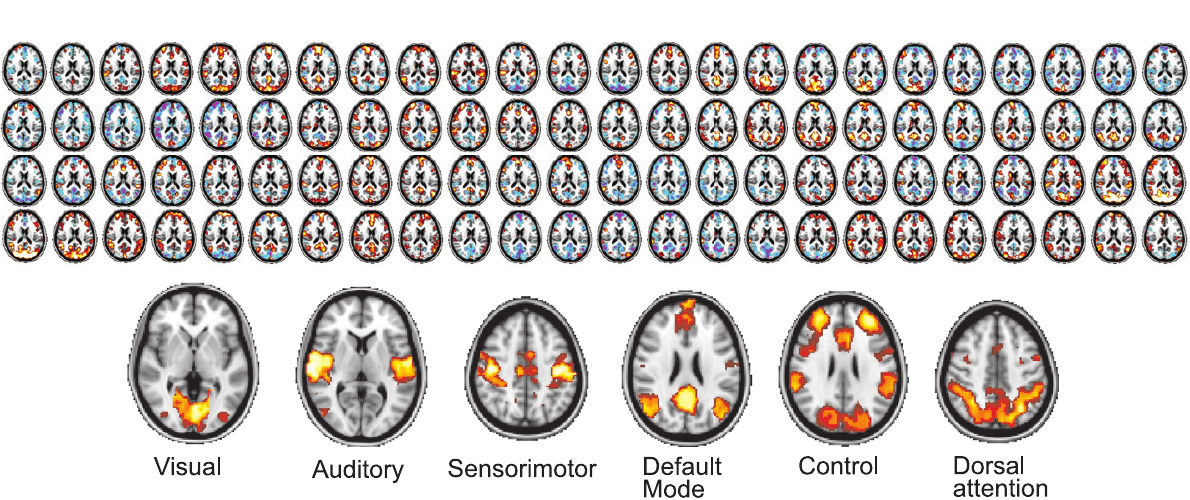}}
\caption{Up: Approximately 10 minutes of resting state brain activity from a single subject (red-yellow for BOLD signal above the mean, blue-purple for BOLD signal below the mean). Images are separated by 2.5 s. Down: Main resting state networks obtained through seed based correlation from data shown above. Replotted from \cite{chialvo2010}.}
\label{dantenature}
\end{figure}

This dynamical scenario is highly consistent with the brain operating in a critical regime. A final observation regarding whole brain measurements is that the lack of measurement scale implied by critical dynamics is also manifest in the invariance of the two-point correlation function of the BOLD signal after a normalization procedure (spatial coarse graining) is repetitively carried out \cite{expert}. Additional evidence is derived from the computation of the correlation length of the BOLD signal fluctuations, which are known to diverge near the critical point. Figure \ref{dani} shows experimental results obtained from human fMRI at rest indicating that correlation length is not constant but diverges with the size of the cluster considered.\cite{chialvofraiman}.  This demonstration is a landmark of critical phenomena, together with previous observations \cite{expert,kitz} of scale invariance at large scale add weigth to the contention of criticality as the brain's dynamical state.
%%%%%%%%%%% Figure 7 %%%%%%%%%%%%%%%%

%%%%%%%%%% FIGURE 8 %%%%%%%%%%%

\section{Neuronal avalanching in small scale is critical}

Historically, the first direct demonstration of collective critical dynamics was the observation that a cultured slice of neurons supported ``avalanches'', this is, intermittent bursts of activity that spread up to the whole system \cite{beggs, plenzTINS}. This dynamical state is halfway between highly synchronized oscillatory activity and disordered noise. Even though pairwise correlations are low, and with highest probably only a small number of neurons fire in synchronized fashion, occasionally activity spreads in the form reminiscent of an avalanche in Per Bak's sandpile model. Power laws are observed for event size (number of electrodes registering electrical activity) with exponents agreeing with those of a critical branching process. These results have been replicated in other experimental settings, including cortical measurements in awake, behaving monkeys \cite{peterman} and rats \cite{sidarta}.

 %%%%%%% FIGURE 9 %%%%%%%%%%%%%%

\begin{figure}[htbp]
\centering{\includegraphics[width=0.5\textwidth]{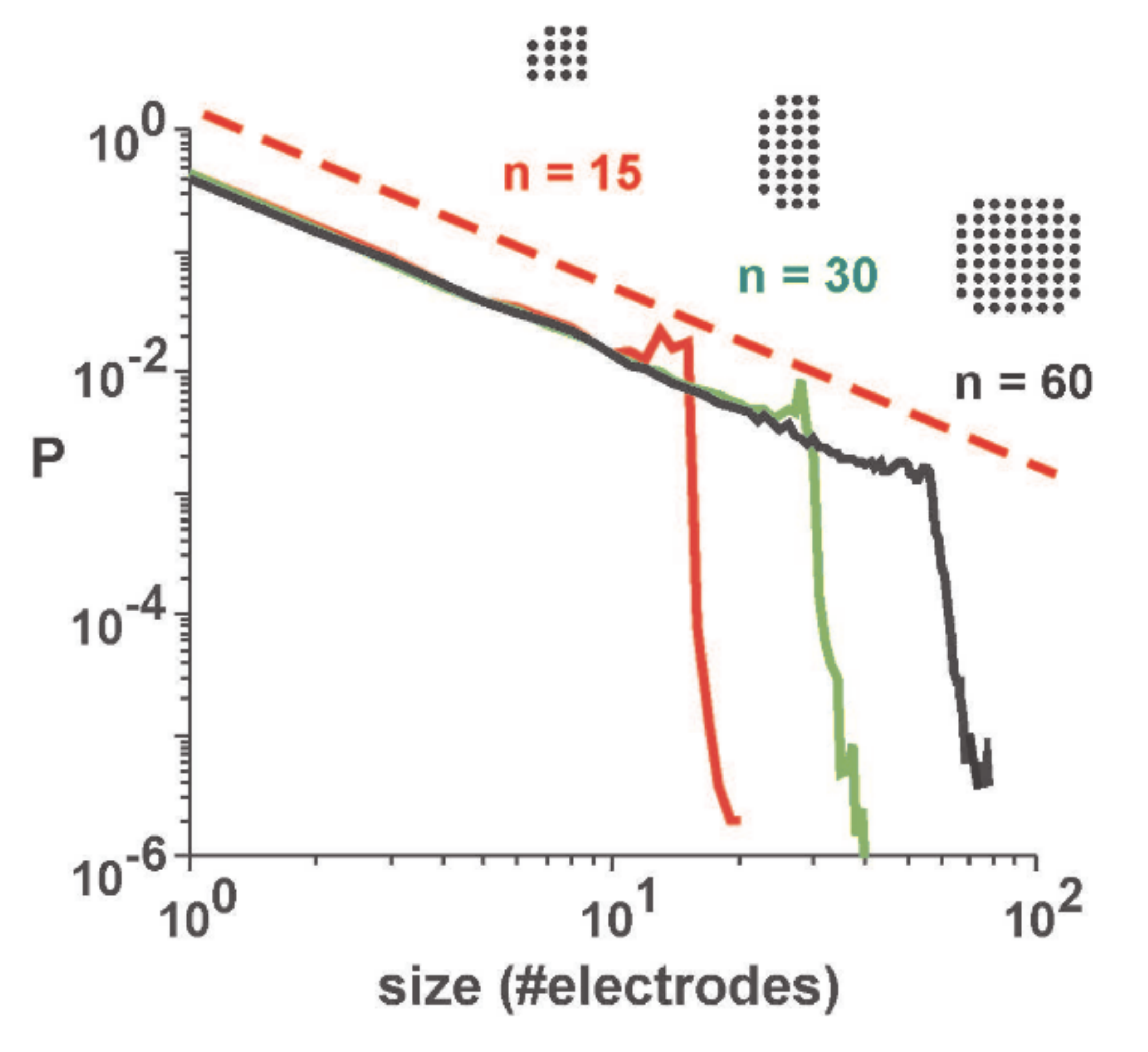}}
\caption{Size distribution for avalanches in mature cortical cultured networks and different electrode array sizes (15,30,60 electrodes). Power laws with different cutoffs arise for each number of electrodes (finite size scaling). Measurements comprise 70 hours of data recording. Replotted from \cite{chialvo2004}.}
\label{avalanches}
\end{figure}

Modeling efforts to understand the precise neuronal cause of avalanches are currently being made. Certain models are able to replicate these scale free densities while  explicitly avoiding criticality, however, recent numerical results show that their predicted exponents are in the wrong order of magnitude as those experimentally observed. Also, even if avalanches are a dynamical process, explanations of scale free behavior can be based on the underlying connectivity of the neurons. However, directly incorporating long range correlations in the structural connectivity of the collectivity solves one problem (integration) but succumbs to a different one (segregation). Critical dynamics is, thus, a natural constraint to incorporate into a model of neuronal avalanches. Models of neurons with activity dependent synaptic coupling undergoing self organized criticality reproduce statistics of neuronal avalanches, even with trivial (random) connectivity of its elements.

Finally, even though the most salient features of experimental findings regarding neuronal avalanches are power laws probability densities for size and event durations, there are other predictions to be fulfilled  if the underlying dynamics are in fact critical. In short, these are: I) time scales separation between the dynamics of the triggering event of the avalanche and the avalanche itself, II) stationary avalanche size statistics regardless of avalanching rate fluctuations (excluding non-homogeneous Poisson processes), III) Omori's law for earthquakes must apply to avalanche probabilities after and before main events, IV) average size of avalanches following a main avalanche decays as an inverse power law, V) avalanches spreads spatially on a fractal. Recently, it has been reported \cite{plenzchialvo} that experimental data in fact supports these predictions, thus narrowing the search of models of avalanches to those undergoing critical dynamics.

\section{Psychophysics and behavior}

As exposed in the previous sections, the hypothesis that collective brain dynamics operates at a critical point has received plenty experimental support. This evidence spans a wide range of scales both in time and space and leads in a natural way to a discussion of consequences at a behavioral level. Simply put: if brain activity is critical, signatures of criticality must be as well observed in behavior and perception.

\begin{figure}[htbp]
\centering{\includegraphics[width=0.6\textwidth]{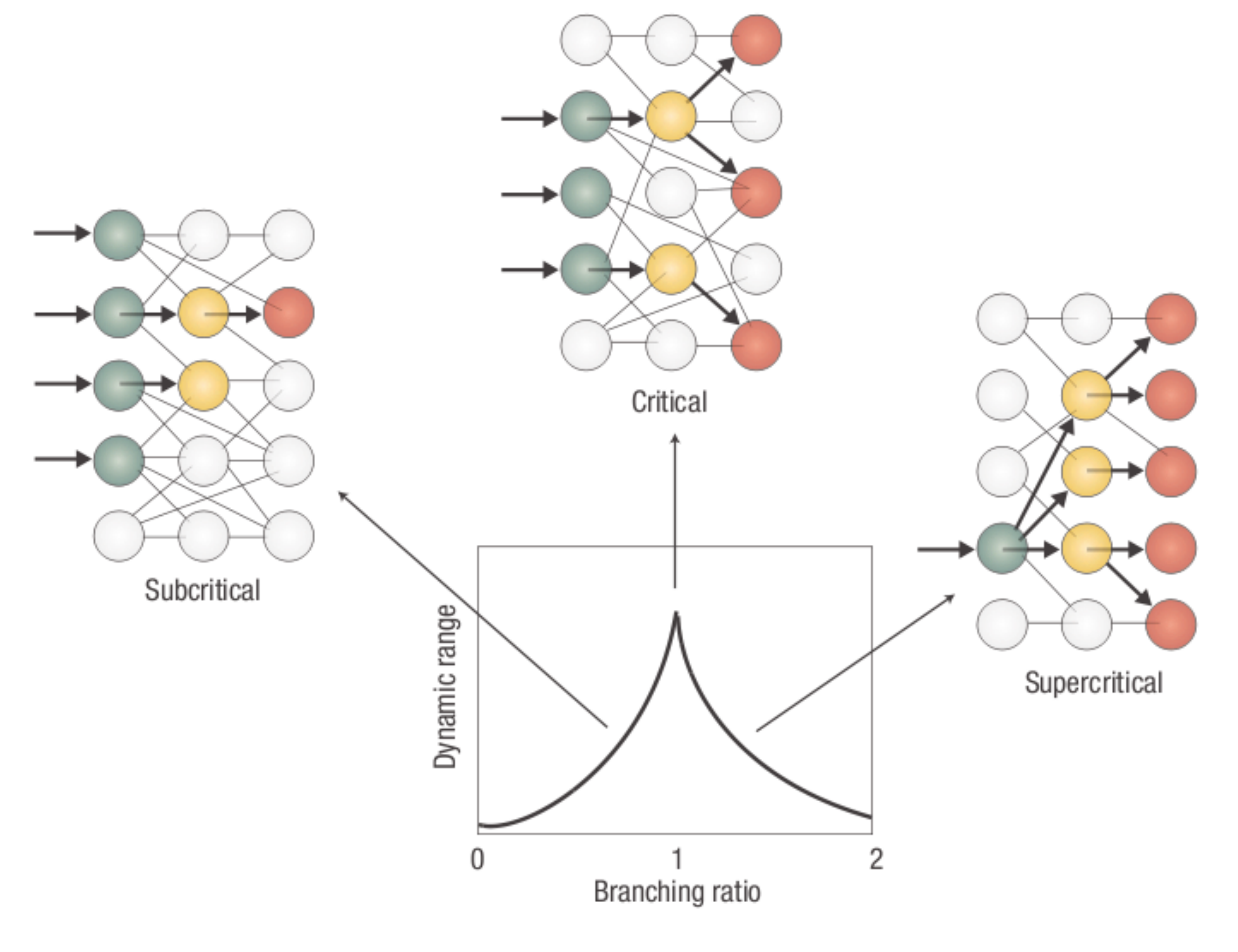}}
\caption{The three dynamical regimes of a sensory network model: subcritical regime (amplification or branching ratio less than one), critical (amplification equal to one), and supercritical (amplification greater than one). In critical networks, activity levels are conserved, and the dynamical range of responses is maximized. Redrawn from \cite{chialvo2006}.}
\label{kopelli}
\end{figure}

An old unsolved problem of psychophysics relates to the need of a very ample dynamic range of neuronal responses, spanning several orders of magnitude\cite{chialvo2006}. For example, visual perception is known to adapt from very dimly lit environments (such as caves, forests, etc)  to landscapes of high luminosity; however, the dynamic range of an isolated neuron (spanning at most a single order of magnitude) is too limited to do so. The key to solve this problem resides in the consideration of a connected network of excitatory neurons which has an amplification factor of one. In this model, neurons operate in a critical point, in contrast to a subcritical regime in which sensory inputs quickly die, or a supercritical regime, in which any input spreads in an explosive fashion throughout the network \cite{kinouchi}. These three possibilities are exemplified in Figure \ref{kopelli}. Thus, the fact that a system at the critical point has the greatest dynamical range, is now the key to solve the dilemma of human perception optimized over several orders of magnitude.

%%%%%%%% Figure 10 %%%%%%%%%%%%%

On the other hand, animal behavior has been shown to reflect the critical dynamics of brain activity. Recordings of spontaneous motion activity of rats during large periods of time reflects scale free distribution for length of movement times and pauses \cite{anteneodo}. This result, observed in an animal model and controlled laboratory conditions, has similarities with reports on human activities such as writing of letters, emails, web browsing and motion initiation. Together, these results suggest the presence of behavioral signatures of critical dynamics in animals and human beings.

\section{An evolutionary perspective}

The ingredients for complex emergent behavior presented in the first section of these notes (a large number of energy-driven nonlinearly interacting elements) are not only present in the human brain, but are also ubiquitous in the whole natural world \cite{bak1,bakbook,maya,buchanan}.
Humans and other animals navigate in this complex critical world, which as such offers constantly surprises and unexpected events. One can argue that in a subcritical and uniform world a brain is a superfluous organ since there is nothing to learn, on the other hand, an ever-changing supercritical world offers no regularities to be learned \cite{bak2, bak3}. In the middle of these two extreme situations there is a real need for a learning  device. The coupling between a critical world and a critical brain is best exemplified in the playing habits of children. Young children become quickly bored of inanimate toys (even the addition of sophisticated motion mechanisms does not dramatically improve this situation), however, they obtain unending fun from animal pets for long periods of time. The reason behind is that animate behavior is a mixture of order and surprise in the same sense than the spatiotemporal patterns of a system at the critical point are composed of a  blend of order and disorder. Thus, the need of a brain stems from the fact that the world around is critical, thus, in turn evolutionary pressures constraint brains to operate in a critical regime. 

Brains must not only be capable of mere learning: the ability to forget is central to adaptive behavior. In a sub-critical brain state of highly correlated activity, memories would be ``frozen'' and  ever present. On the other hand, a supercritical brain would have its activity patterns changing wildly and constantly in time, resulting in the inability to retain any memories. The conclusion is that, in order to adapt to a complex world that constantly presents us with novelties, evolution has created brains and forced them operate at the critical point.

\section{Noise or critical fluctuations? Equilibrium vs non-equilibrium.}

As discussed in Section II, criticality is a universal dynamics exhibited by a large class of systems. Consequently it offers a unified theoretical framework to explain a myriad of apparently unrelated observations.  We will briefly provide an illustration of this possibility, discussing work recently reported, where the current theoretical understanding can be fruitfully applied. 

In a series of papers \cite{misic,garret} the variance and mean-based spatial fMRI patterns obtained in subjects of different  ages were compared, concluding that the descriptive value of the variance was greater than the averages. The authors suggested that \emph{``examination of BOLD signal variability may reveal a host of novel brain-related effects not previously considered in neuroimaging research"}. Additional work (by the same authors) went even further, revealing that BOLD signal variability is greater at younger age, confirming that \emph{ ``younger, faster, and more consistent performers exhibited significantly higher brain variability across tasks, and showed greater variability-based regional differentiation compared to older, poorer performing adults"}\cite{garret2}.

The origin of this variability was explained \cite{misic} by suggesting that:
\emph{``The interplay between local and global dynamics governs the
spatiotemporal configuration of the brain's functional architecture
and keeps the system in a high-energy state, at the `edge of instability' among a number of different states and configurations."}, while the switch between such configurations was suggested to happen by the action of noise originated in the following manner: \emph{``Intrinsic neural noise --stochastic fluctuations in information transfer caused by imprecise timing of cellular processes-- serves to nudge the system from one state to another and thus confers the capacity to make fluid and adaptive transitions between different states and reconfigure either spontaneously or in response to external (task) demand." }

Besides the obvious merits of calling attention to the variability,  the crucial question here should be how much real understanding was gained by finding that the variability was more informative than the average. \footnote{The first and second moments are usually linked by some functional relation, thus it is not completely surprising that averages and variance share predictability.} But lets postpone the answer to this question until we consider yet another example.

The second example is concerned with a very interesting book \cite{rollsdeco} dedicated recently to the topic of noise in the brain, which in its very first five lines reads: \emph{``The relatively random spiking times of individual neurons produce a source of noise in the brain. The aim of this book is to consider the effects of this and other noise on brain processing. We show that in cortical networks this noise is an advantage, for it leads to probabilistic behavior that is advantageous in decision making, by preventing deadlock, and is important in signal detectability"}.  This is not an isolated quote taken out of the proper context. The authors enumerate a three ``reasons why the brain is inherently noisy and stochastic"\cite{rollsdeco} including sensory noise arising external to the brain, cellular noise due to stochastic opening and closing of ion channels, and synaptic noise. 

These two examples  are in line with nearly all detailed models of neuronal function which require ad-hoc noise inputs to function properly. Simply put, without noise the system is at equilibrium, stuck on a stable state. All these analysis overlook a crucial question: where does this (fine tuned or not) noise comes from? If it is generated by specific systems within the brain, then experimental evidence (``a noisy center" ?) should be found pointing to where these systems are and how they work. Of course, sarcasms aside, no such center would be found. Instead, ad-hoc noisy driving is not required by the theory of critical brain dynamics, since noise is  self-generated by the collective dynamics which spontaneously fluctuates near the critical point as was shown in Figure 3.  It is well known, that large fluctuations are inherent to a system approaching the critical point \cite{bakbook}, and in that sense ``variability" is perfectly understood as part of the critical dynamics, and not a force external to the system.  In our opinion, the often conjectured reasons for the noise presence, assigning a purpose to it \footnote{``the noise serves to...,"}, are \emph{teleological} dead-ends that postpone true understanding.
 
The fact is that the correct interpretation of the noise is a much deeper issue, which reflects the general tendency to consider only equilibrium models. These models achieve a better agreement with empirical observations only with the addition of \emph{external} noise (see for instance Chapter 12 in \cite{sporns2010} or work compiled by Steyn-Ross \cite{steyn} recently). This deformation is not exclusive of neuroscience and its consequences are far reaching. For example, similar views in other complex systems include the explanation of episodes with very large number of species going extinct throughout earth history by adding external noise, such that for each extinction an out of space event (meteorite) hits the earth. In contrast, non-equilibrium models \cite{bakbook,maya} show that similar exceptionally rare and large extinctions (and their mirror situations, namely speciations) can be inherent of the co-evolutionary species dynamics, and that large extinctions could have happened in absence of meteorites. In these type of models, variability and disparate fluctuations are (at criticality) typical. Another example belongs to macro-economics, as the late Per Bak hammered many times in his writings\cite{bakbook} and lectures, criticizing models in which each market crash would have required an event that drove the shares to plunge. In analogy, historians will assure us that in Sarajevo on the 28th of June of 1914 a member of the Black Hand, a Serbian nationalist secret society, assassinated  Archduke Franz Ferdinand, heir to the Austro-Hungarian throne. The equilibrium view will identify this single event (noise?) as responsible to trigger the mobilization of 65 millions soldiers, 7 millions missing, 8 millions killed and 21 millions wounded. In contrast, the non-equilibrium approach will assert that the First War World (and other infrequently large conflicts) will eventually happen in absence of an isolated gunman.
 
Thus, coming back to the brain, where its main dynamics  are fluctuations, it need to be recognized that having to add fluctuations (noise) to the model equations to get fluctuating dynamics is a serious limitation. It seems reasonable that future work consider that the differences between modeling a ``noisy brain" or a ``critical brain" are significant and relevant,  because the gap separating these two views is conceptually as large as for physics is the difference between equilibrium and non-equilibriums dynamics. 

\section{Outlook}
The study of collective dynamics and critical phenomena was pioneered by physicists during the past century, in order to solve problems related with material science and solid state physics. It is now clear that the theoretical framework used for these problems is very general and can be applied to understand an impressive range of phenomenology from many disciplines. Emergent dynamical properties of interacting agents (phase transitions, divergence of correlation length and fluctuations, maximum dynamical range, etc) are far from explicit in the individual equations of motion, thus, a reductionist approach will be (by its very definition) unable to explain this emergence.

Pure reductionism has in fact been abandoned in a large list of fields which study complex many particle systems, from economics and sociology to molecular biology. It is surprising that this revolution has only recently began to have an impact on neuroscience, a discipline which deals more than any other with a collective of interacting agents. Still today, a large number of published results stem from detailed experiments or simulations of neurons or networks of millions of neurons, including sometimes great biological detail but no considerations related with the statistical physics of collective phenomena in large systems. However, after the realization that theories of brain function must be compatible with the laws of collective behavior, a non-return point has been reached: a rapidly increasing number of reports being deal with theoretical and experimental evidence grounding the emergent critical dynamics hypothesis \cite{chialvo2010}.

 Researchers now find that paradoxical situations and seemingly disconnected facts can be re conciliated within the critical brain dynamics theory, without need of complicated ad-hoc hypothesis. A good example of this is the integration/segregation dilemma \cite{tononi98a, tononi98, tononi2004}, a problem haunting neuroscientists since the beginnings of the discipline: how does the brain manages to integrate information from different modalities into a single decomposable scene, while making easy to segregate particular aspects of this scene at will? Or in other terms: how does brain activity from different regions coordinates without collapsing into a single block of uniform dynamics? A concrete example is the synchronous coexistence of multiple operational modules studied by Fingelkurts \& Fingelkurts \cite{fingelkurts}.  Many connectionistic approaches have been taken to this problem, implying  that coordination is hardwired into anatomical connectivity. These proposed solutions are able to solve one problem (integration) but still fail to explain the ease of brain dynamics to segregate. Within the theory of critical brain dynamics, however, this situation arises naturally: even systems with trivial structural connectivity (for example, the first-neighbor connections of the Ising model) achieve (transient) long range correlations at the critical point \cite{fraiman}.

Any scientific theory needs to explain and integrate previous facts, solve apparently paradoxical situations, and make predictions amenable to experimental verification. Along these notes we have shown how these requirements are fulfill by the theory of emergent brain collective critical dynamics. The transfer of ideas from statistical physics to other disciplines is a young and exciting endeavor, even younger in neuroscience\cite{werner}, however exciting results have already been achieved. Only with the realization that brain is a collective (and henceforth must follow the physical laws that govern collectives) a deep understanding of brain and human behavior will be achieved.

\section*{Acknowledgements}Work supported by NIH (USA)  and by CONICET (Argentina).  E.T. was supported by an Est\'imulo Fellowship from the University of Buenos Aires.

%%%%%%%%%%%%%%%%%%%%%%%%%%%Sectional units are obtained with the \LaTeX{} commands:

%\bibliographystyle{ws-procs9x6}
%\bibliography{ws-pro-sample}

\end{document}